\documentclass[12pt]{article}

\usepackage{epsfig,amsmath,amssymb,verbatim,mathrsfs}

\numberwithin{equation}{section}

\def\beq{\begin{eqnarray}}
\def\eeq{\end{eqnarray}}
\def\bea{\begin{eqnarray}}
\def\eea{\end{eqnarray}}

\def\gev{\, {\rm GeV}}

\newcommand{\gsim}{\lower.7ex\hbox{$\;\stackrel{\textstyle>}{\sim}\;$}}
\newcommand{\lsim}{\lower.7ex\hbox{$\;\stackrel{\textstyle<}{\sim}\;$}}

\def\stilde{\widetilde}

\newcommand{\newc}{\newcommand}
\newc{\Nc}{N_{c}}
\newc{\CG}{C_G}
\newc{\gp}{g'}
\newc{\stopi}{\stilde t_i}
\newc{\sboti}{\stilde b_i}
\newc{\staui}{\stilde \tau_i}
\newc{\stopj}{\stilde t_j}
\newc{\sbotj}{\stilde b_j}
\newc{\stauj}{\stilde \tau_j}
\newc{\stopI}{\stilde t_1}
\newc{\stopII}{\stilde t_2}
\newc{\sbotI}{\stilde b_1}
\newc{\sbotII}{\stilde b_2}
\newc{\stauI}{\stilde \tau_1}
\newc{\stauII}{\stilde \tau_2}
\newc{\sstop}{s_{t}}
\newc{\cstop}{c_{t}}
\newc{\ssbot}{s_{b}}
\newc{\csbot}{c_{b}}
\newc{\sstau}{s_{\tau}}
\newc{\cstau}{c_{\tau}}
\newc{\Sstop}{s_{2t}}
\newc{\Cstop}{c_{2t}}
\newc{\Ssbot}{s_{2b}}
\newc{\Csbot}{c_{2b}}
\newc{\Sstau}{s_{2\tau}}
\newc{\Cstau}{c_{2\tau}}
\newc{\salpha}{s_\alpha}
\newc{\calpha}{c_\alpha}
\newc{\Calpha}{c_{2\alpha}}
\newc{\Salpha}{s_{2\alpha}}
\newc{\sbetapm}{s_{\beta_\pm}}
\newc{\cbetapm}{c_{\beta_\pm}}
\newc{\Sbetapm}{s_{2 \beta_\pm}}
\newc{\Cbetapm}{c_{2 \beta_\pm}}
\newc{\sbetaO}{s_{\beta_0}}
\newc{\cbetaO}{c_{\beta_0}}
\newc{\SbetaO}{s_{2 \beta_0}}
\newc{\CbetaO}{c_{2 \beta_0}}
\newc{\vu}{v_u}
\newc{\vd}{v_d}
\newc{\seL}{\stilde e_L}
\newc{\smuL}{\stilde \mu_L}
\newc{\seR}{\stilde e_R}
\newc{\smuR}{\stilde \mu_R}
\newc{\suL}{\stilde u_L}
\newc{\sdL}{\stilde d_L}
\newc{\suR}{\stilde u_R}
\newc{\sdR}{\stilde d_R}
\newc{\scL}{\stilde c_L}
\newc{\ssL}{\stilde s_L}
\newc{\scR}{\stilde c_R}
\newc{\ssR}{\stilde s_R}
\newc{\snue}{\stilde \nu_e}
\newc{\snumu}{\stilde \nu_\mu}
\newc{\snutau}{\stilde \nu_\tau}
\newc{\Gpm}{G^\pm}
\newc{\Hpm}{H^\pm}
\newc{\FFbS}{\overline{FF}S}
\newc{\FFbV}{\overline{FF}V}
\newc{\FSS}{F_{SS}}
\newc{\FSSS}{F_{SSS}}
\newc{\FFFS}{F_{FFS}}
\newc{\FFFbS}{F_{\overline{FF}S}}
\newc{\FSSV}{F_{SSV}}
\newc{\FVS}{F_{VS}}
\newc{\FVVS}{F_{VVS}}
\newc{\FFFV}{F_{FFV}}
\newc{\FFFbV}{F_{\overline{FF}V}}
\newc{\Fgauge}{F_{\rm gauge}}
\newc{\DRbarprime}{$\overline{\rm DR}'$ }
\newc{\DRbar}{$\overline{\rm DR}$ }
\newc{\MSbar}{$\overline{\rm MS}$ }
\newc{\Yu}{{\bf Y}_u}
\newc{\Yd}{{\bf Y}_d}
\newc{\Ye}{{\bf Y}_e}
\newc{\Au}{{\bf a}_u}
\newc{\Ad}{{\bf a}_d}
\newc{\Ae}{{\bf a}_e}
\newc{\bm}{{\bf m}}
\newc{\zhol}{Z^{\rm hol}}

\newc{\rwino}{r_{\tilde W}}
\newc{\rmu}{r_{\tilde H}}
\newc{\ra}{r_A}
\newc{\ccdot}{\!\cdot\!}

\newcommand{\nnmb}{\nonumber}
\newcommand{\del}{\partial}

\newcommand{\lrf}[2]{\left(\frac{#1}{#2}\right)}

\newcommand{\ttt}{\tilde{t}}
\newcommand{\tel}{\tilde{\ell}}

\newcommand{\gmu}{\Gamma\,G\mu}

\begin{document}

\setlength{\baselineskip}{0.2in}



\begin{titlepage}
\noindent
\begin{flushright}
MCTP-08-09\\
NSF-KITP-08-17
\end{flushright}
\vspace{1cm}

\begin{center}
  \begin{Large}
    \begin{bf}
Non-Thermal Dark Matter from Cosmic Strings\\
     \end{bf}
  \end{Large}
\end{center}
\vspace{0.2cm}

\begin{center}

\begin{large}
Yanou Cui$^{a,b}$ and David E. Morrissey$^{a}$\\
\end{large}
\vspace{0.3cm}
  \begin{it}
$^a$~Michigan Center for Theoretical Physics (MCTP) \\
Physics Department, University of Michigan, Ann Arbor, MI 48109
\vspace{0.2cm}\\
$^b$~Kavli Institute for Theoretical Physics \\
University of California, Santa Barbara, CA 93106-4030
\vspace{0.5cm}
\\
  \vspace{0.3cm}
\vspace{0.1cm}
\end{it}

\end{center}

\center{\today}

\begin{abstract}

  Cosmic strings can be created in the early universe during
symmetry-breaking phase transitions, such as might arise if the
gauge structure of the standard model is extended by additional
$U(1)$ factors at high energies. Cosmic strings present in the
early universe form a network of long horizon-length segments, as
well as a population of closed string loops.  The closed loops are
unstable against decay, and can be a source of non-thermal
particle production.  In this work we compute the density of WIMP
dark matter formed by the decay of gauge theory cosmic string
loops derived from a network of long strings in the scaling
regime or under the influence of frictional forces. We find that 
for symmetry breaking scales larger than 
$10^{10}\,\gev$, this mechanism has the potential to 
account for the observed relic density of dark matter.  
For symmetry breaking scales lower than this, the density of 
dark matter created by loop decays from a scaling string network 
lies below the observed value.  In particular, the cosmic 
strings originating from a $U(1)$ gauge symmetry broken 
near the electroweak scale, that could lead to a massive $Z'$ 
gauge boson observable at the LHC, produce a negligibly small 
dark matter relic density by this mechanism.

\end{abstract}

\vspace{1cm}

\end{titlepage}

\setcounter{page}{2}


\vfill\eject



\section{Introduction}

  Cosmic strings are one-dimensional topological
defects~\cite{Nielsen:1973cs,Hindmarsh:1994re,sbook}. They can be
formed during symmetry breaking phase transitions in the early
universe~\cite{Kibble:1976sj}, such as might arise if the gauge
symmetry group of the standard model~(SM) is enlarged by
additional $U(1)$ factors at high energies~\cite{abelian,muterm,gut}.
Such gauge factors appear naturally in many models of physics beyond the SM.
A typical cosmic string network consists of horizon-length
\emph{long strings} as well as smaller closed \emph{string loops}.
The long strings carry a net conserved topological charge and are stable,
while closed string loops do not have a net charge and can decay away.
In the present work we study the production of dark matter by the
decays of cosmic string loops~\cite{Jeannerot:1999yn,Matsuda:2005fb}.

  The evolution of cosmic string networks in the early universe
has been studied extensively.  It is found that the long strings
in the network stretch with the Hubble expansion, as well as form
closed string loops by the process of \emph{reconnection} when
they intersect themselves or each
other. After an initial transient, these two processes balance
out such that the string network reaches a \emph{scaling}
regime, in which the total energy density of the network makes up
a small, constant fraction of the dominant matter or radiation
energy density of the universe~\cite{Kibble:1984hp,Bennett:1987vf,
Allen:1990tv,Martins:1995tg,Vincent:1996rb,Vanchurin:2005pa}. This
string energy fraction is largely independent of the initial
conditions, and is about $G\mu$ relative to the critical density,
where $G = 1/(8\pi M_{\rm Pl}^2)$ is Newton's constant and $\mu$
is the string tension.  In terms of the scale of spontaneous
symmetry breaking $\eta$ from which the strings arise, the tension
is on the order of 
\beq 
\mu \simeq \eta^2. 
\eeq 
Cosmic string scaling can be modified if the strings have significant
interactions with the thermal background that lead to an effective
frictional force on the strings.  The importance of friction to
the evolution of a string network depends on the pattern of
symmetry breaking as well as the temperature of the surrounding
plasma.

  A possible decay product of the string loops formed by a scaling
cosmic string network is dark matter~(DM)~\cite{Jungman:1995df}. Indirect
evidence for cold dark matter has been obtained from a number of
sources. Together, they indicate a dark matter density
of~\cite{Komatsu:2008hk} \beq \Omega_{DM}h^2 = 0.1143\pm0.0034
\eeq This dark matter density can be explained by the existence of
a new weakly-interacting stable particle with a mass on the order
of the electroweak scale~(WIMP)~\cite{Jungman:1995df}.
Such particles are common in
extensions of the SM that stabilize the electroweak scale against
quantum corrections such as supersymmetry~\cite{Martin:1997ns},
universal extra dimensions~\cite{Appelquist:2000nn}, and little
Higgs models with $T$-parity~\cite{Cheng:2004yc}.

   String loops can produce DM and other particles
in a number of ways.  For local cosmic strings, corresponding to
the spontaneous breakdown of a gauge symmetry, the direct emission
of particles by the strings is thought to be
suppressed~\cite{Srednicki:1986xg}.\footnote{However, see
Refs.~\cite{Vincent:1996rb,Borsanyi:2007wm} for arguments to the
contrary.}  Instead, loops lose most of their energy by
oscillating and emitting gravitational
radiation~\cite{Vilenkin:1981bx}. However, when the loop radius
shrinks down to the order of the string width, it will
self-annihilate into its constituent
fields~\cite{Jeannerot:1999yn}. The subsequent decays of these
states can produce dark matter. An even larger source of string
annihilation and particle production by closed loops is the formation of
\emph{cusps}~\cite{Brandenberger:1986vj}. These are points on a
loop where the string segment folds back on itself and briefly
approaches the speed of light~\cite{Kibble:1982cb,Turok:1984cn}.
In the vicinity of a cusp, a small portion of the string will
self-annihilate creating particles~\cite{Brandenberger:1986vj,
BlancoPillado:1998bv}.  In many simple solutions for the motion of
a string loop, a cusp is generally found to occur about once per
loop oscillation period~\cite{Kibble:1982cb,Turok:1984cn,
Vachaspati:1984gt,Burden:1985md,Garfinkle:1987yw,Quashnock:1990wv}.

  The amount of dark matter created by a cosmic string network
depends on the initial size of the string loops that are formed by
the network.  While the evolution of the long horizon length
strings is well understood, the details of loop formation are less
clear. These details are closely related to the spectrum of small
fluctuations on long strings. Significant advances have been made
recently in this direction, both in numerical
simulations~\cite{Vanchurin:2005yb,Ringeval:2005kr,
Martins:2005es}, as well as in analytic
models~Ref.~\cite{Siemens:2002dj,
Polchinski:2006ee,Polchinski:2007qc,Rocha:2007ni,Dubath:2007mf,
Vanchurin:2007ee}.  In the present work we will mostly adopt the
results of Refs.~\cite{Polchinski:2006ee,
Polchinski:2007qc,Rocha:2007ni,Dubath:2007mf} to characterize the
initial loop size spectrum.

  In the picture of loop formation (in the scaling regime) that emerges from
Refs.~\cite{Polchinski:2006ee,Polchinski:2007qc,Rocha:2007ni,Dubath:2007mf},
fluctuations on long strings are created near the horizon scale
$d_H\sim t$, which is also the scale that characterizes the long
string network in the scaling regime.  After they are formed, these
fluctuations grow less quickly than the horizon, and thereby shrink
relative to the characteristic scale of the long string network.
The fluctuation spectrum that emerges is a power law in the fluctuation
size that increases going to smaller scales.  This power law is eventually
cut off well below the horizon by gravitational radiation damping,
which erases very small fluctuations.  The cut-off occurs when the
fluctuation size falls below the gravitational radiation
scale~\cite{Polchinski:2006ee}
\beq
l_{GW} = \Gamma\,(G\mu)^{1+2\,\chi}\,t
\label{lgw}
\eeq
where $\Gamma \simeq 50$ is a constant, $t$ is the cosmic time,
and $\chi = 0.10$ during
radiation and $0.25$ during the matter era.
The small fluctuation spectrum on long strings is therefore
peaked near $l_{GW}$.  This peak implies that $l_{GW}$ sets the
typical initial loop size $\ell_i$~\cite{
Polchinski:2006ee,Polchinski:2007qc,Rocha:2007ni,Dubath:2007mf},
\beq
\ell_i \simeq l_{GW}.
\eeq
Both the number and the energy density of the loops formed are
dominated by loops of this initial size.\footnote{For smaller 
values of $\eta$ and at very early times, the gravitational 
damping length $l_{GW}$ 
can fall below the width of the string, $w\sim \eta^{-1}$.
In this case, we expect that small fluctuations are cut off 
at $\ell \sim w > l_{GW}$ by direct particle emission as suggested 
in Ref.~\cite{Vincent:1996rb,Borsanyi:2007wm}.}
However, recent simulations (and the analytic model of
Ref.~\cite{Vanchurin:2007ee}) also point toward a significant loop
population near the horizon scale, with $\ell_i\simeq (0.1)\,t_i$~\cite{
Vanchurin:2005yb,Ringeval:2005kr,Martins:2005es}.
We will therefore consider larger loops as well in our analysis.

  The primary goal of this paper is to compute the dark matter
density generated by a network of gauge theory cosmic strings in the scaling
regime. We also calculate the dark matter generated when
\emph{frictional} interactions of the strings with the background 
plasma modify the evolution of the network.  A necessary condition for the
phenomenological viability of a cosmic string network is that 
the dark matter density it generates not exceed the observational bound.  
Our results therefore provide a constraint on theories of new physics
that lead to the formation of cosmic strings, such as models
containing new $U(1)$ gauge groups that arise frequently in
superstring theory constructions and in models of grand
unification~\cite{abelian,muterm,gut}.  Our results also 
motivate certain classes of gauge extensions of the standard model
that are able to generate the observed dark matter relic density
by cosmic string loop decay.
While we concentrate on the dark matter 
produced by a network of cosmic strings, our
methods can also be applied to computing the densities of other
cosmologically interesting particles arising from cosmic string loop
decays, such as moduli and gravitinos.

  We find that for symmetry breaking scales $\eta$ larger 
than $10^{10}\,\gev$, the decays of cosmic string loops 
derived from a network of long cosmic strings that are scaling or
dominated by friction can potentially generate (more than) 
enough cold dark matter to account for the observed relic density.
The precise density depends on the typical initial loop size and
the effective branching fraction of the loop decays into cold dark matter,
in addition to $\eta$.  Note that $\eta = 10^{10}\,\gev$
corresponds to $G\mu \simeq 10^{-18}$, for which the standard
cosmic string signatures such as gravitational radiation and
gravitational lensing are expected to be unobservably 
weak~\cite{Hindmarsh:1994re,sbook}. 
When $\eta$ is much smaller than $10^{10}\,\gev$,
including those scales for which the massive $Z'$ gauge boson
associated with the symmetry breaking could be visible at the LHC,
the density of non-thermal dark matter produced by  
this mechanism is well below the observed value.  Thus, such models 
are not constrained by dark matter overproduction from 
scaling string loops.

  This paper is organized as follows.  In Section~\ref{formula}
we derive a general formula for the production of dark matter by
cosmic string loops.  In Section~\ref{scaling}, we apply this
formula to compute the amount of dark matter produced by string
loops in the scaling regime.  In Section~\ref{friction} we extend our
results to compute the dark matter density from cosmic string
loops when friction plays an important role in the evolution of
the string network. Finally, Section~\ref{conclusion} is reserved
for our conclusions.

  Before proceeding, let us point out that the production of dark matter
by the decays of cosmic string loops was considered previously for
\emph{ordinary} Abelian Higgs model
cosmic strings in Ref.~\cite{Jeannerot:1999yn}, for cosmic strings
derived from a supersymmetric flat direction in Ref.~\cite{Cui:2007js},
and for cosmic strings associated with aspects of superstring
theory in Refs.~\cite{Matsuda:2005fb,Damour:1996pv,Peloso:2002rx}.
We concentrate on ordinary cosmic strings in the present work, and we
update and extend the results of Ref.~\cite{Jeannerot:1999yn} in a
number of ways.  Most importantly, we focus on cusping as the
primary source of particle production by cosmic strings.
We also make use of the recent results on the size
distribution of string loops when they are formed from Refs.~\cite{
Polchinski:2006ee,Polchinski:2007qc,Rocha:2007ni,Dubath:2007mf},
and we consider the evolution of cosmic string networks both with
and without friction.  When friction is relevant, we extend
Ref.~\cite{Jeannerot:1999yn} by using the analytic model of
Ref.~\cite{Martins:1995tg} to describe the long string network.

\section{A Formula for Dark Matter from Cosmic Strings\label{formula}}

  Let us denote the initial \emph{invariant} length of a string
loop formed at cosmic time $t_i$ by $\ell_i$.
The invariant length $\ell$ of a loop is defined in relation to its
energy in the cosmological frame by
\beq
E_{loop} = \mu\,\ell.
\eeq
If the loop is boosted with speed $\nu$, the invariant length $\ell$
will exceed the proper length of the loop in its rest frame
by a factor of $\gamma = 1/\sqrt{1-\nu^2}$.

  The key quantity characterizing loop formation is
\beq r(\ell_i,t_i)\,d\ell_i\,dt_i, \eeq the number density of
string loops formed in the time interval $(t_i,t_i+dt_i)$ with
initial length in the range $(\ell_i,\ell_i+d\ell_i)$. The form of
the function $r(\ell_i,t_i)$ is constrained by the evolution of
the long string network.  In particular, the total rate at which
energy is transferred from the long string network to loops is
typically a known quantity~\cite{Martins:1995tg}. We can use this
fact to impose the constraint
\beq
\frac{d\rho_{loop}}{d t_i} = \int
d\ell_i\;\mu\,\ell_i\,r(\ell_i,t_i), \label{rnorm}
\eeq
where $d\rho_{loop}/dt$ is the rate of energy
transfer into loops from the long string network.

  Consider the evolution of a loop of initial size $\ell_i$
formed at time $t_i$.  The length of this loop is
described by the function
\beq
\ell(t;\ell_i,t_i)
\eeq
which is the solution to the equation
\beq
\mu\frac{d\ell}{dt} = -P_{tot},~~~~~\mbox{with}~~~~~
\ell(t_i;\ell_i,t_i) = \ell_i,
\eeq
where $P_{tot}$ is the total rate of energy loss from the loop.
We will assume that $P_{tot}$ is positive so that $\ell$ is monotonically
decreasing in time.
Under this assumption, the loop will eventually decay away completely
at the time $t_{co}(\ell_i,t_i)$, defined implicitly by
\beq
0 = \ell(t_{co};\ell_i,t_i).
\eeq
While the loop is decaying, it will emit a fraction of its energy
in the form of (cold) dark matter.  We will write this fraction as
$P_{DM} \leq P_{tot}$.

  To find the total rate of dark matter production by the collection
of string loops, consider loops with initial size in the range
$(\ell_i,\ell_i+d\ell_i)$ formed in the time interval
$(t_i,t_i+dt_i)$. There are $r(\ell_i,t_i)\,d\ell_i\,dt_i$ such loops
per unit volume initially. As time goes on, this collection of
loops is diluted by the cosmic expansion by a factor of $a^{-3}$.
This is the \emph{only} modification of their density until they
decay away completely at time $t_{co}(\ell_i,t_i)$. At time
$\ttt$, this collection of loops will produce dark matter at the
rate $P_{DM}$.  The dark matter will thermalize if it is produced
before the freeze-out time $\ttt < t_{fo}$, and redshift as
$a^{-3}$ after this time.  Thus, we have
\beq
\Delta\rho_{DM} =
\int d\ell_i\int_{t_{\eta}}^{t_0} dt_i\,
\int_{t_{fo}}^{\bar{t}_{co}(\ell_i,t_i)}
\!\!\!d\ttt\;\;r(\ell_i,t_i)\left[\frac{a(t_i)}{a(\ttt)}\right]^3\,
P_{DM}\,\left[\frac{a(\ttt)}{a(t_0)}\right]^3
\;\Theta(t_{co}-t_{fo}). \label{mastereq1}
\eeq
The integral over $t_i$ runs between the string network
formation time $t_{\eta}$ and the present time
$t_0 \simeq 4\times 10^{41}\,\gev^{-1}$.
Typically, $t_{\eta}$ is on the order of
\beq
t_{\eta}
\simeq \frac{M_{\rm Pl}}{\eta^2},
\eeq
where $\eta$ is the scale of
spontaneous symmetry breaking. The integration over the decay time
$\ttt$ runs from the freeze-out time $t_{fo}$ to either $t_{co}$
or $t_0$, whichever is smaller. Thus, we have defined
$\bar{t}_{co}(\ell_i,t_i)$ by
\beq
\bar{t}_{co}(\ell_i,t_i) =
\left\{
\begin{array}{ll}
t_{co}(\ell_i,t_i)&:~~~t_{co}< t_0\\
t_0&:~~~t_{co}\geq t_0
\end{array}
\right..
\eeq
Any restriction on the integration limits of $\ell_i$ are encoded
in the support of the function $r(\ell_i,t_i)$.

  It is often the case that $P_{DM}$ and $P_{tot}$ depend on
$\ttt$ only through $\tel = \ell(\ttt;\ell_i,t_i)$.
If so, it is convenient to change the
integration variable from $\ttt$ to $\tel$.  The corresponding
Jacobian is simply $(\del\tel/\del\ttt)^{-1} = -\mu/P_{tot}$,
where $P_{tot}$ is the total power released by a loop of length
$\tel$. It follows that
\beq
\Delta \rho_{DM} = \int
d\ell_i\int_{t_{\eta}}^{t_0} dt_i\,
\int_{\ell_x(\ell_i,t_i)}^{\bar{\ell}_{fo}(\ell_i,t_i)}d\tel\;
r(\ell_i,t_i)\left[\frac{a(t_i)}{a(t_0)}\right]^3\,
\mu\,\frac{P_{DM}}{P_{tot}}
\label{mastereq2}
\eeq
The limits on the $\tel$ integration depend on the functions
\beq
\bar{\ell}_{fo}(\ell_i,t_i) = \left\{
\begin{array}{ll}
  \ell(t_{fo};\ell_i,t_i)&:~~~t_i< t_{fo}\\
  \ell_i&:~~~t_i> t_{fo}
\end{array}
\right.,
\eeq
as well as
\beq
\ell_x(\ell_i,t_i) = \left\{
\begin{array}{ll}
  0&:~~~t_{co}(\ell_i,t_i) < t_{0}\\
  \ell(t_0;\ell_i,t_i)&:~~~t_{co}(\ell_i,t_i) \geq t_{0}
\end{array}
\right..
\eeq
Eqs.~\eqref{mastereq1} and \eqref{mastereq2} are our main results.
We will apply them to two interesting special cases below.

\section{Dark Matter Production from a Scaling Network\label{scaling}}

  As a first application of our main result, Eq.~\eqref{mastereq1},
we compute the dark matter density produced by
a network of cosmic strings in the scaling regime due to the
cusping of closed string loops.  To simplify the analysis,
we focus on monochromatic loop formation distributions with
\beq
\ell_i = \alpha\,t_i.
\eeq
This relation implies $r(\ell_i,t_i) \propto \delta(\ell_i-\alpha t_i)$.
More general distributions can be obtained by introducing a weight function
and summing the final result over different values of $\alpha$.

  In the string scaling regime, the rate at which the long string
network transfers energy into loops is~\cite{Martins:1995tg}
\beq
\frac{d\rho_{loop}}{dt_i} = \zeta\,\mu\,t_i^{-3},
\label{loopprod1}
\eeq
with $\zeta \simeq 10$ a constant characterizing the mean properties
of the long string network.  Imposing the constraint of
Eq.~\eqref{rnorm} on the loop formation rate $r(\ell_i,t_i)$, we obtain
\beq
r(\ell_i,t_i) = \frac{\zeta}{\alpha}\,t_i^{-4}\,\delta(\ell_i-\alpha\,t_i).
\label{rmono}
\eeq
All that remains to do is to specify the evolution of the loop length
and the power emitted as dark matter, and apply Eq.~\eqref{mastereq1}.

  Cosmic string loops lose energy to gravitational radiation
as well as cusping, and shrink as a result.  The rate of energy emission
into gravity waves is~\cite{sbook}
\beq
P_{grav} = \Gamma\,G\mu^2,
\label{pgrav}
\eeq
where $\mu$ is the string tension, $G = 1/(8\pi\,M_{\rm Pl}^2)$,
and $\Gamma \simeq 50$ is a dimensionless constant~\cite{Vachaspati:1984gt}.

  Loops also lose energy when they form cusps, which are points
on the loops that briefly fold back upon themselves and approach
the speed of light.  Near the apex of a cusp, a portion of the string
overlaps itself and self-annihilates.  Summing over many cusps,
the net rate at which a loop loses
energy to cusping is given by~\cite{BlancoPillado:1998bv}
\beq
P_{cusp} = \mu\,p_c\,\sqrt{\frac{w}{\ell}},
\label{pcusp}
\eeq
where $w\sim \eta^{-1}$ is the string width, and
$p_c$ is the probability for a cusp to form per period
of loop oscillation.  Several studies suggest $p_c\simeq 1$~\cite{
Turok:1984cn,Vachaspati:1984gt,Burden:1985md,Garfinkle:1987yw,
Quashnock:1990wv}.
Comparing the cusp power of Eq.~\eqref{pcusp} to the gravitational wave
power of Eq.~\eqref{pgrav}, we see that cusping is the dominant energy-loss
mechanism by loops when they are shorter than $\ell < \ell_=$ with
\beq
\ell_= = w\,\lrf{p_c}{\Gamma G\mu}^2.
\label{leq}
\eeq
The particles produced by cusping can decay into dark matter.
Provided $p_c \gtrsim \Gamma G\mu$, we expect cusping to be
the dominant source of particle production and dark matter from cosmic strings.

  If loops lose energy to gravitational radiation and
cusping, the loop length evolves according to
\beq
\mu\frac{d\ell}{dt} = -P_{tot} = -\Gamma G\mu^2 - \mu\,p_c\,
\sqrt{\frac{w}{\ell}}.
\label{ptot}
\eeq
The solution of this equation subject to the initial condition
$\ell(t_i) = \ell_i$ is given implicitly by
\beq
t-t_i = \frac{\ell_=}{\Gamma G\mu}\left[
\lrf{\ell_i-\ell}{\ell_=}
- 2\left(\sqrt{\frac{\ell_i}{\ell_=}}-\sqrt{\frac{\ell_i}{\ell_=}}\right)
+2\,\ln\lrf{1+\sqrt{\ell_i/\ell_=}}{1+\sqrt{\ell/\ell_=}}\right].
\label{tofl}
\eeq

  Suppose that a fraction $\epsilon_{cusp}$ of the energy
emitted by a string loop cusp (eventually) takes the form of cold dark matter.
It follows that
\beq
P_{DM} = \epsilon_{cusp}\,P_{cusp}.
\eeq
Putting this into Eq.~\eqref{mastereq2} and making use of Eq.~\eqref{rmono},
we find the dark matter density due to cusping to be
\bea
\Delta\rho_{DM} &=& \int_0^{\infty}d\ell_i\int_{t_{\eta}}^{t_0}dt_i\,
\int_{\ell_x(\ell_i,t_i)}^{\ell_{fo}(\ell_i,t_i)}d\tel\;
r(\ell_i,t_i)\,\lrf{a_i}{a_0}^3\,\mu\,
\frac{\epsilon_{cusp}P_{cusp}(\tel)}{P_{tot}(\tel)}
\label{cuspdm}\\
&=& \int_{t_{\eta}}^{t_0}dt_i\;\frac{\zeta}{\alpha}\,t_i^{-4}\,
\lrf{a_i}{a_0}^3\,2\,\epsilon_{cusp}\,\mu\,
\ell_=\,\left[\sqrt{\frac{\ell}{\ell_=}} -
\ln\left(1+\sqrt{{\ell}/{\ell_=}}\right)
\right]_{\ell_x(\alpha\,t_i,t_i)}^{\bar{\ell}_{fo}(\alpha\,t_i,t_i)}\;
\Theta(\bar{\ell}_{fo})\,\Theta(\ell_x).\nnmb
\eea
The $\Theta$ functions in this expression account for the fact
that only those loops that decay after $t_{fo}$ and before $t_0$
can contribute to the dark matter density.  Numerically, we find
that the integrand of Eq.~\eqref{cuspdm} is a steeply falling 
function of $t_i$ that is cut off at small $t_i$ by the 
$\Theta(\bar{\ell}_{fo})$ condition.  
It follows that the dominant contribution to the dark matter 
density comes from loops formed at the smallest possible
value of $t_i$ such that they decay shortly after $t_{fo}$.
In this sense, the dark matter density is created nearly instantaneously
at $t_{fo}$. 

  To be concrete, we evaluate Eq.~\eqref{cuspdm} assuming
a weakly-interacting massive dark matter particle with a freeze-out
time of $t_{fo} = 2\times 10^{16}\,\gev^{-1}$.  This corresponds
approximately to a freeze-out temperature of $5\,\gev$, which is
in the range expected for a stable, weakly-interacting particle
with a mass on the order of $100\,\gev$.  
We take the universe to be radiation dominated prior to $t_{fo}$, 
and we assume a standard cosmological evolution afterward.
We also set the loop network parameter to $\zeta = 10$, 
the cusping probability to $p_c=1$,
and the branching fraction into dark matter equal to unity,
$\epsilon_{cusp} = 1$.  In realistic models $\epsilon_{cusp}$
can be much smaller than unity.  Our results can therefore be
interpreted as providing an upper bound on $\epsilon_{cusp}$ within
the underlying gauge theory model.

\begin{figure}[ttt]
\begin{center}
        \includegraphics[width = 0.7\textwidth]{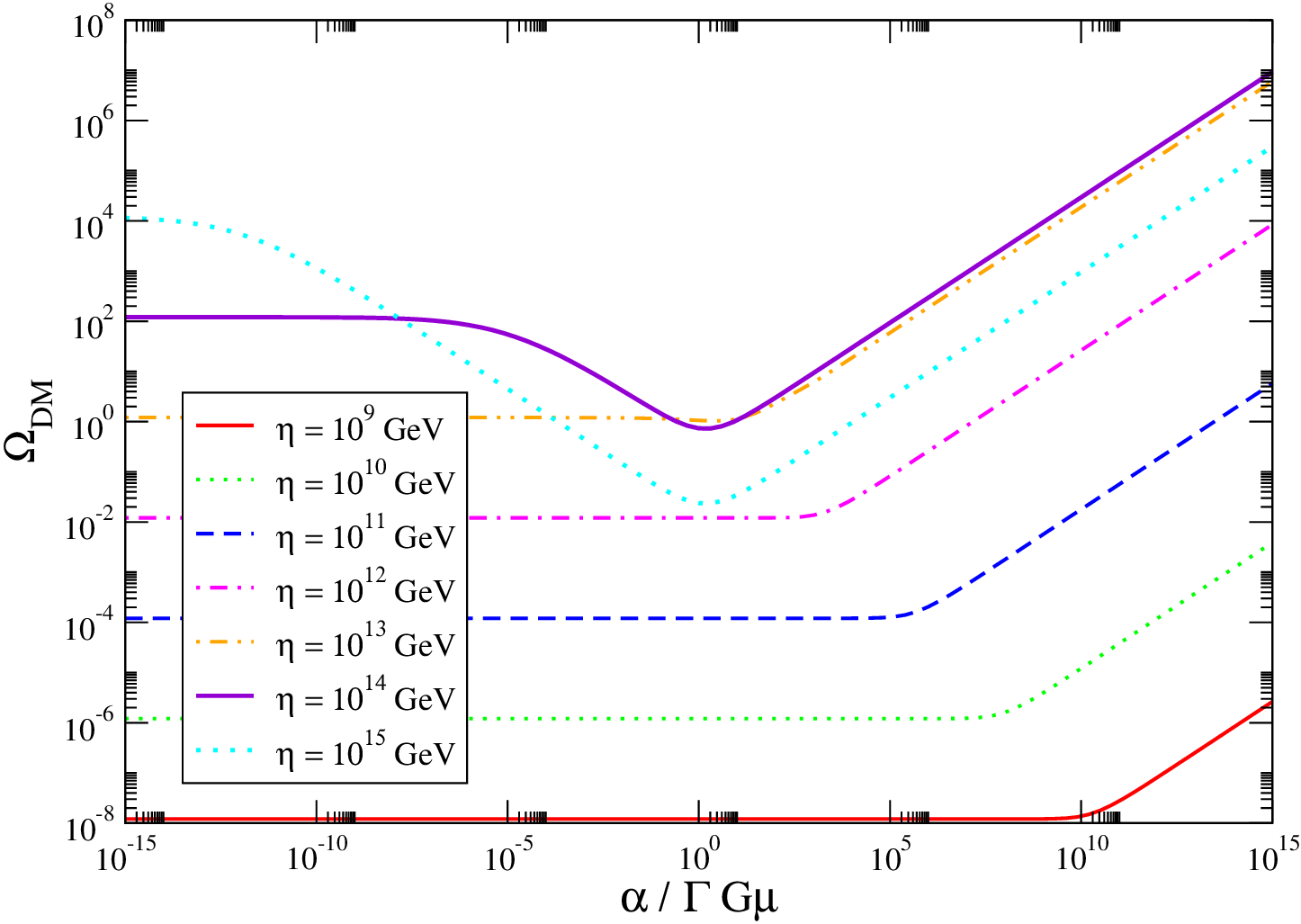}
\end{center}
\vspace{0.2cm} \caption{Dark matter density due to loop cusping
for $\epsilon_{cusp} = 1$, $p_c=1$, $\zeta = 10$, and $t_{fo} =
2\times 10^{16}\,\gev^{-1}$ as a function of the initial loop size
parameter $\alpha$. The various lines correspond to different
values of the symmetry breaking VEV $\eta$.} \label{cuspalf}
\end{figure}

  In Fig.~\ref{cuspalf} we show the dark matter relic density due
to cusping as a function of the initial loop size parameter
$\alpha$, normalized to $\Gamma G\mu$, with the other parameter
values as described above.  We normalize $\alpha$ to $\Gamma G\mu$
because, in the absence of cusping, a loop is long-lived relative
to the Hubble time at formation ($\sim t_i$) provided
$\alpha/\Gamma G\mu > 1$, and short-lived otherwise.
From this plot, we see that increasing the initial loop size $\alpha$
well above $\Gamma G\mu$ increases the final dark matter density.
The reason for this is that the integration over $t_i$ in
Eq.~\eqref{cuspdm} is dominated by the earliest times for which
loops produced at $t_i$ decay after $t_{fo}$.  Since larger loops are
longer-lived, loops contributing to the dark matter density can be
formed at earlier times when the integrand of Eq.~\eqref{cuspdm} is larger.
As $\alpha/\Gamma G\mu$ decreases, the dark matter density curves
flatten out for lower values of the VEV $\eta$.  In these flat regions,
cusping dominates the evolution of the loops contributing to the dark matter,
and these loops are short-lived even though $\alpha/\Gamma G\mu > 1$.
When the VEV is larger, $\eta \gtrsim 10^{13}\,\gev$,
gravitational radiation becomes more important than cusping.
Loops then become short-lived only when $\alpha/\gmu < 1$.
When $\alpha$ becomes very small, cusping again takes over and
the curves flatten out once more.

\begin{figure}[ttt]
\begin{center}
        \includegraphics[width = 0.7\textwidth]{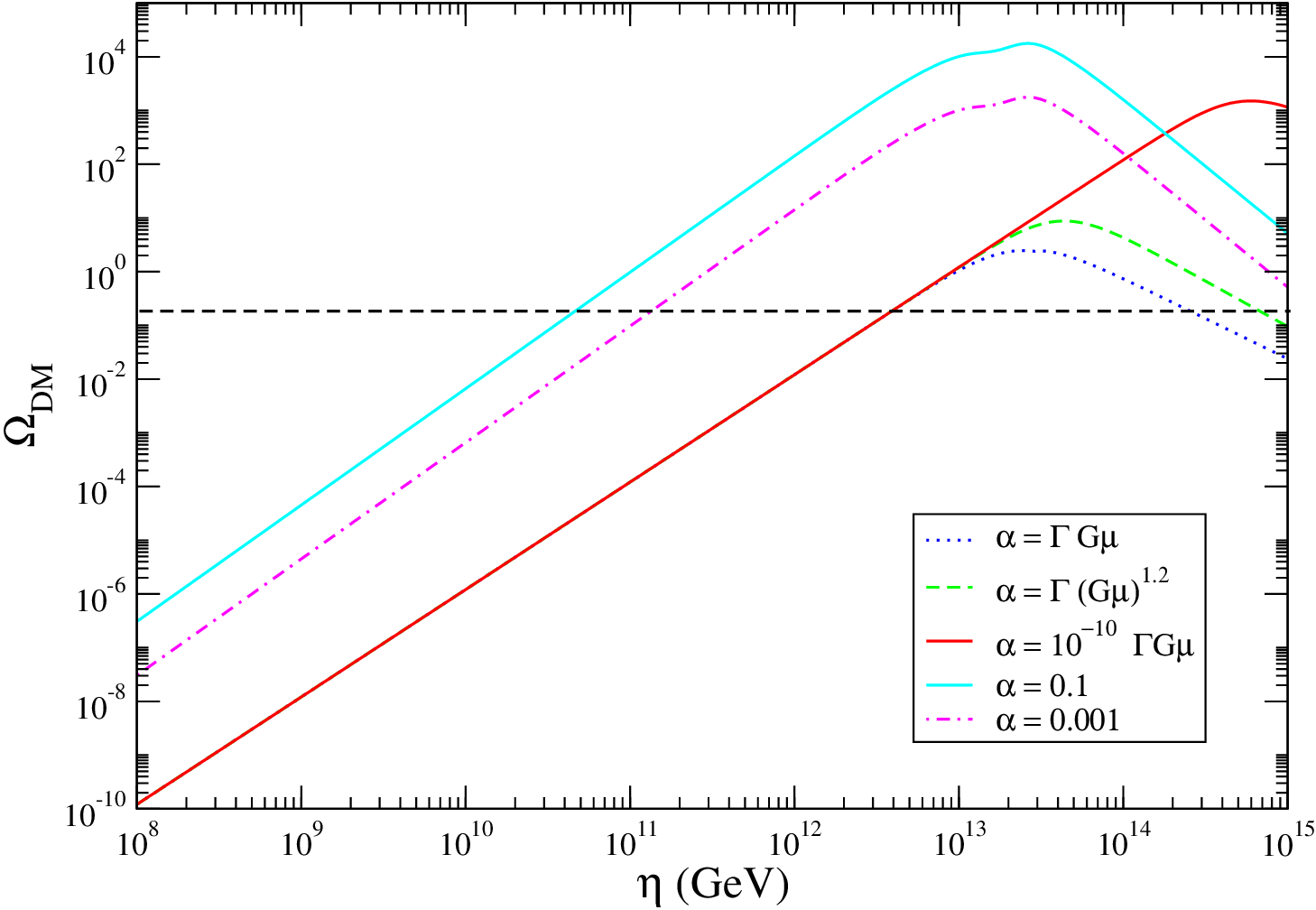}
\end{center}
\vspace{0.2cm}
\caption{Dark matter density due to loop cusping for $\epsilon_{cusp} = 1$,
$\zeta = 10$, $p_c=1$, and $t_{fo} = 2\times 10^{16}\,\gev^{-1}$
as a function of the symmetry breaking VEV $\eta$.  The various curves
correspond to different values of the initial loop size parameter $\alpha$.
The black dashed horizontal line denotes the observed dark matter
relic density.
}
\label{cuspvev}
\end{figure}

 In Fig.~\ref{cuspvev} we show the dependence of the dark matter
density from string cusping as a function of the symmetry breaking
VEV $\eta$. The other string model parameters are as described
above. This plot shows a general increase in the dark matter
density up to large values of $\eta$, and then a fall-off.
Short-lived loops, with $\alpha \leq \gmu$, and smaller values of
$\eta$ decay mostly through cusping at time $t_{fo}$. Using
Eq.~\eqref{mastereq2}, it is possible to show that $\Omega_{DM}$
increases as $\eta^2$ in this case. As $\eta$ is increased
further, gravitational radiation becomes more important than
cusping, and loops lose most of their energy to gravity waves
instead of dark matter. Thus, the dark matter density falls for
these very large values of $\eta$, with $\Omega_{DM} \propto
(\alpha\,\eta)^{-1/2}$ in this regime. The precise value of $\eta$
at which the cross-over occurs depends on the value of $\alpha$,
with smaller initial loops being more prone to cusping. For very
large initial loops, $\alpha =0.1$, a similar cross-over occurs.

\begin{figure}[ttt]
\begin{center}
        \includegraphics[width = 0.7\textwidth]{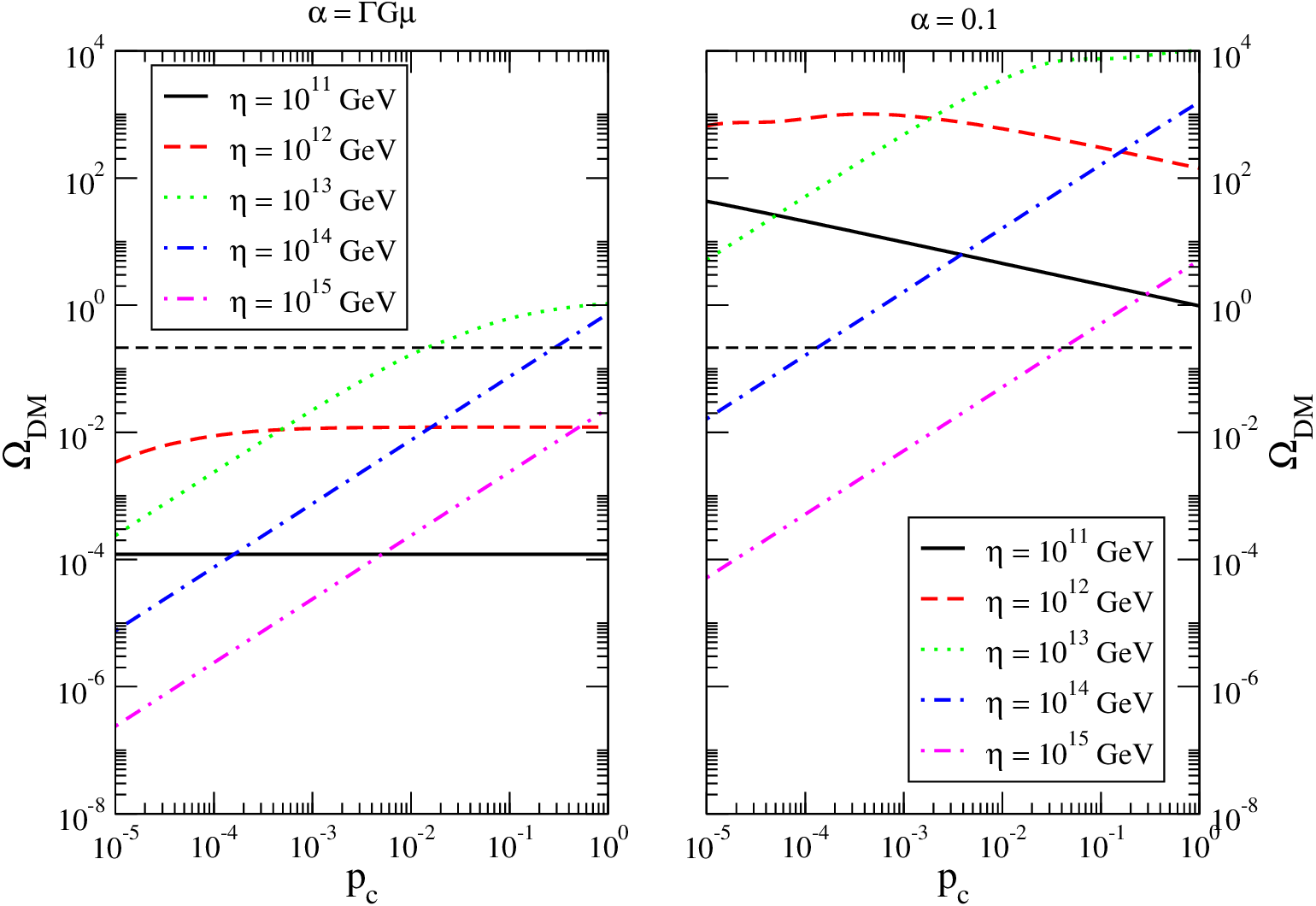}
\end{center}
\vspace{0.2cm}
\caption{Dark matter density due to loop cusping for $\epsilon_{cusp} = 1$,
$\zeta = 10$, and $t_{fo} = 2\times 10^{16}\,\gev^{-1}$
as a function of the cusp formation probability $p_c$ for several
values of the symmetry breaking scale $\eta$.
The panel on the left corresponds to small initial loops
with $\alpha = \Gamma G\mu$, while the panel on the right corresponds
to large initial loops with $\alpha = 0.1$.
The black dashed horizontal line indicates the observed dark matter
relic density.
}
\label{cuspcc}
\end{figure}

  In general, we expect the loop cusping probability $p_c$ to be
on the order of unity.\footnote{The backreaction from string annihilation
at a cusp can suppress its reoccurrence, but this need not prevent the
formation of new cusps~\cite{BlancoPillado:1998bv}.}
However, it is also interesting to look at
smaller values of $p_c$, such as might arise on large string loops 
with many kinks~\cite{Garfinkle:1987yw,Quashnock:1990wv}.
Fig.~\ref{cuspcc} shows the effect of reducing $p_c$ for small 
($\alpha = \gmu$) and
large ($\alpha = 0.1$) initial loop sizes, and a range of values of the
symmetry breaking VEV $\eta$. All other parameters are as in the
previous plots. For $\alpha = \gmu$ the loops are necessarily
short-lived. In this case, cusping is the dominant energy loss mechanism 
at time $t_{fo}$ for $\eta \lesssim 10^{12}\,\gev$, and remains so
even for lower values of $p_c$. Thus, reducing $p_c$ does not
alter the resulting dark matter density.
When $\eta$ is larger, gravitational radiation dominates the 
energy loss at time $t_{fo}$, and reducing $p_c$ therefore 
decreases the fraction of loop energy released as dark matter. 
With $\alpha = 0.1$ string loops can be long-lived. For smaller 
values of $\eta \lesssim 10^{12}\,\gev$, 
lowering $p_c$ can enhance the dark matter density.  
This occurs because cusping is the dominant energy loss mechanism 
around time $t_{fo}$ for these loops. Reducing $p_c$ thus further 
increases the lifetime of these long-lived loops which, as discussed above,
enhances the final dark matter density. For larger values of
$\eta$, gravitational radiation dominates at time $t_{fo}$, so
that reducing $p_c$ only decreases the fraction of energy released
as dark matter without significantly altering the loop lifetime.

  We have only computed here the contribution to the dark matter density
from loops formed in the scaling regime.  Loops created in the
string-forming phase transition at time $t_{\eta}$ and while
the string network was evolving towards scaling will give an additional
contribution to the dark matter density if they decay after $t_{fo}$.
The approach of a string network to scaling depends on the details of the phase
transition, and a study of this process is beyond the scope of the present
work.  However, we can derive the condition for loops formed in the
phase transition to decay before $t_{fo}$.  We obtain
\beq
\eta \gtrsim
p_c^{-2/5}\,\lrf{\alpha}{0.1}^{3/5}\,\left(7\times 10^3\,\gev\right).
\eeq
If this bound is satisfied and scaling is attained quickly after
the network is formed at $t_{\eta}$, loops formed in the scaling regime
will give the dominant contribution to the dark matter density.
This bound also indicates that our results are applicable to the
phenomenologically interesting case of a $U(1)$ gauge symmetry
broken only slightly above the electroweak scale
provided $p_c \sim 1$ and $\alpha \ll 0.1$.

  Before moving on, let us mention two additional effects that
we have not yet taken into account.  The first is that we have treated 
the loops as having no net motion relative to the thermal background,
even though it is found in Refs.~\cite{Polchinski:2006ee,
Polchinski:2007qc,Rocha:2007ni,Dubath:2007mf} that small loops
($\alpha < \Gamma\,G\mu$) can be highly boosted.  This boost
does not alter the cusping power, although the expression for this
power in Eq.~\eqref{pcusp} picks up a factor of 
$\gamma^{1/2}$ (for short-lived loops),
where $\gamma = 1/\sqrt{1-\nu^2}$ is the boost factor.
The decay products will also be boosted.  Unless the loops are 
very tiny, however, this boost is not expected to greatly modify 
our results.

  A second and potentially much larger effect that we have 
not yet included is that the decay products from a loop cusp can 
be very highly boosted, even if the loop is initially at rest.  
The size of this boost $\gamma$ was estimated in 
Ref.~\cite{BlancoPillado:1998bv} to be
\beq
\gamma = \sqrt{\frac{\ell}{w}}.
\eeq
On account of this boost, the cusp decay products including the dark matter
might not thermalize immediately upon production.  If not, they may
thermalize at a later time, or even remain as hot dark matter.

  To estimate the effect of this possibly very large boost, 
we adopt a simple picture of the cusp decay products as
consisting of a \emph{bunch} of light states (including WIMP dark matter)
moving together with a boost $\gamma$ relative to the thermal background.
In the rest frame of the bunch we assume very little relative motion,
and that a fraction $\epsilon_1$ of the total bunch energy consists of
dark matter particles of mass $m \simeq 100\,\gev$.  The dark matter in
the bunch will thermalize when $\sigma\,n_{tot} \gtrsim H$, 
where $n_{tot}\sim T^3$ is the total plasma number density of light states 
and $\sigma$ is the cross-section for the dark matter to scatter
off the thermal background.  To estimate the thermalization time, 
we take this cross-section to be
\beq
\sigma = \frac{1}{8\pi}\,\frac{1}{s} = \frac{1}{8\pi}\,\frac{1}{\gamma^2m^2}.
\eeq
With this cross-section, the dark matter decay products from a 
cusp on a small loop with $\alpha =\Gamma G\mu$ formed at time 
$t_{fo}$ will reach kinetic equilibrium nearly instantaneously after the 
cusp occurs when the symmetry breaking scale lies below 
$\eta \lesssim 10^{12}\,\gev$.  In this case, our previous estimates 
for the dark matter density remain valid.
  
  If the cusp decay products do not thermalize instantaneously,
they may still thermalize at a later time $t_T$ as their momentum redshifts.  
To compute the final density of cold dark matter
in this case, it is only necessary to add a factor of $a(\tilde{t})/a(t_T)$ 
to the integrand of Eq.~\eqref{mastereq1} and modify the limits
of integration such that $t_T > t_{fo}$ rather than $\tilde{t} > t_{fo}$.
Doing so, we find that the resulting dark matter density coincides with
what we found previously (without the boost) up to an overall factor
of order unity even when instantaneous thermalization does not occur.  
The additional redshift $a(\tilde{t})/a(t_T)$ is cancelled nearly
exactly by the delay in termalization, which allows loops contributing
to the dark matter density to be formed at earlier times when the rate
of loop production was larger.  Similar to before, we also find that
the dominant contribution to the cold dark matter density comes from
loop decay products that thermalize shortly after $t_{fo}$.  

  The prospect of highly boosted decay products also raises several
new issues.  Since a large fraction of the energy of the cusp typically
goes into the kinetic energy of the decay products, the fraction
of this energy that ends up as \emph{cold} dark matter can be
suppressed.  If kinetic thermalization proceeds exclusively through
elastic collisions, we would have $\epsilon_{cusp} = \epsilon_1/\gamma(t_T)$,
where $\gamma(t_T)$ is the boost factor at the time of thermalization.
The actual value of $\epsilon_{cusp}$ can be larger than this if inelastic
collisions during thermalization generate additional dark matter
particles, but in general we expect a significant suppression
of $\epsilon_{cusp}$ relative to the non-boosted case.  
Highly boosted loop decay products can also generate relics that 
have not thermalized by the present time (or the time of matter-radiation 
equality).  In general, the resulting densities of hot dark matter 
are suppressed by factors of $G\mu$ and are safely small for lower 
values of the symmetry breaking scale.  However, a reliable precise estimate 
is complicated by the fact that for extremely large boosts, 
the cusp decay products will na\"ively have super-Planckian momenta.  
A full analysis of this issue is beyond the scope of the present work.

  In summary, our results indicate that the density of WIMP cold 
dark matter generated by string loops from cusping with 
$p_c = 1$ in the scaling regime is safely below the observed value for 
$\eta \lesssim 10^{10}\,\gev$ ($G\mu\simeq 10^{-18}$).
For values of the symmetry-breaking scale larger than this,
the string-induced dark matter density depends on the initial loop size.
Loops of initial size $\alpha = \Gamma\,(G\mu)^{1.2}$, motivated by the results
of Refs.~\cite{Polchinski:2006ee,Polchinski:2007qc,Rocha:2007ni,Dubath:2007mf},
can create as much or more than the observed dark matter density
for $\eta \gtrsim 5\times 10^{12}\,\gev$ ($G\mu\simeq 10^{-13}$)
if the branching fraction $\epsilon_{cusp}$ is on the order of unity.
Larger loops, with $\alpha \sim 0.1$, can lead to a much greater
dark matter density.  If such large loops are typical, the branching 
fraction into dark matter $\epsilon_{cusp}$ must be significantly 
less than unity if the cosmic strings are to avoid overproducing 
dark matter.  This is natural for large loops, whose cusps generate
very highly boosted decay products, leading to much of the cusp
energy being lost to kinetic energy that is transferred to 
the thermal bath rather than cold dark matter.
Let us also emphasize that for models with a new $U(1)$ gauge 
symmetry broken only slightly above the electroweak scale, 
the cold dark matter produced by the corresponding cosmic
string network in the scaling regime is negligibly small 
for both large and small initial loop sizes.

\section{Dark Matter Production with Friction\label{friction}}

  The evolution of a cosmic string network can be modified if the
strings have significant interactions with the thermal background.
As we will discuss below, the relevance of these interactions to
the string network depends on the details of the symmetry breaking
from which the cosmic strings arise. Such interactions, when
present, tend to slow the motion of the strings by creating an
effective frictional force on them~\cite{
Vilenkin:1991zk,Garriga:1993gj,Martins:1995tg}. This in turn
changes the density and rate of growth of the network. The
frictional forces on strings decrease as the universe cools, and
eventually become unimportant relevant to the Hubble damping from
the expansion of spacetime.  Frictional effects also change the
way cosmic string loops form and decay, and can enhance
the total rate of loop formation. In this
section we apply the result of Eq.~\eqref{mastereq1} to compute
the density of dark matter created by a cosmic string network
evolving under the influence of frictional forces.

  For local (gauge) cosmic strings, the dominant interaction between
the strings and the thermal background comes from Aharonov-Bohm
scattering~\cite{Alford:1988sj}. This results from the phase
change experienced by the charged particle as it is transported
around the string. The effective frictional force induced by this
scattering can be characterized by a \emph{friction length}
$\ell_f$.  Friction becomes unimportant when $\ell_f(t)$ grows larger
than the Hubble length.  The friction length due to Aharonov-Bohm
scattering is given by~\cite{Vilenkin:1991zk,Martins:1995tg}
\beq
\ell_f = \frac{\mu}{\beta\,T^3}, \label{lfric}
\eeq
where the dimensionless quantity $\beta$ is~\cite{Vilenkin:1991zk}
\beq
\beta = \frac{2\,\zeta(3)}{\pi^2}\,\sum_a\,b_a\,\sin^2(\pi\nu_a),
\eeq
with the sum running over relativistic degrees of freedom,
and $b_a = 1\,(3/4)$ for bosons (fermions).  The value of
$\nu_a$ is related to the charge $Q_a$ of the light particle species $a$.
If the underlying $U(1)$ gauge symmetry (subgroup) is broken by
the condensation of a field with charge $Q$, it is equal to $Q_a/Q$.
Therefore we expect $\beta$ to be of order unity when some
of the $\nu_a$ are non-integer, and zero otherwise.
When Aharonov-Bohm scattering vanishes,
\emph{Everett scattering} of charged particles off the strings
will be the dominant source of frictional interactions~\cite{Everett:1981nj}.
The corresponding friction length is similar to that
for (non-trivial) Aharonov-Bohm scattering but is enhanced by a factor of
$\ln^2({T}/{\eta})$.  Frictional interactions will be
completely irrelevant when all the light states in the theory
are uncharged under the broken gauge group.  In the present section,
we will assume that Aharonov-Bohm scattering is the dominant
source of frictional interactions with $\beta = 1$.

  Frictional effects on the long string network decouple
when the friction length grows larger than the Hubble
length. This occurs at the time $t_*$ defined by the relation 
\beq
H(t_*) = 1/\ell_f(t_*). 
\eeq 
Friction is only relevant for the
long string network when $t<t_*$. For $\beta = \mathcal{O}(1)$ and
radiation domination, $t_*$ has the parametric size\footnote{ Due
to the many uncertainties involved in the description of cosmic
strings within the friction-dominated regime, we only list and use
here the leading parametric dependences of the string network
parameters.} 
\beq 
t_* \simeq \frac{M_{\rm Pl}^3}{\eta^4},
\label{tstar} 
\eeq 
where $\eta$ denotes the symmetry breaking VEV.
This is parametrically larger than the typical (radiation-era)
formation time $t_{\eta} \sim M_{\rm Pl}/\eta^2$.  

  The evolution of a cosmic string network in the presence of friction
was studied in Ref.~\cite{Martins:1995tg}.  In their analytic
model, the long string network is characterized by an effective
correlation length $L$ and a mean velocity $\nu$. The energy
density of the network is given in terms of these variables as
\beq 
\rho_{\infty} = \frac{\mu}{L^2}. 
\eeq 
If the initial string
density is larger than the scaling density, as would be expected
if the symmetry breaking phase transition is second-order or
weakly first-order, the string network
evolves very quickly to the \emph{Kibble
regime}~\cite{Hindmarsh:1994re,Martins:1995tg}. In this regime,
with the universe assumed to be radiation dominated, the string
network variables $L$ and $\nu$ have the parametric
dependences~\cite{Martins:1995tg} 
\bea
L(t) &\simeq& \lrf{t}{t_*}^{1/4}\,t\label{lkib}\\
\nu(t) &\simeq& \lrf{t}{t_*}^{1/4}.\label{vkib}
\eea
The Kibble regime only lasts while $t < t_*$.  From Eqs.~\eqref{lfric}
and \eqref{tstar} we see that the friction length grows as
\beq
\ell_f \simeq \lrf{t}{t_*}^{1/2}t.
\label{lfric1}
\eeq
As $t$ approaches $t_*$, the friction length catches up to the
long string length $L$ as well as the horizon,
and the string network transitions into the usual
scaling regime with $L\propto t$ and $\nu\sim 1$.

  During the friction-dominated Kibble regime, the rate at
which energy is transferred from the long string network to loops
is on the order of
\beq
\frac{d\rho_{loop}}{d t_i} \simeq
\mu\,\lrf{t_*}{t_i}^{1/2}\,t_i^{-3},~~~~~t_i < t_*.
\label{loopprod2}
\eeq
This is parametrically larger than during the scaling regime,
as can be seen by comparing with Eq.~\eqref{loopprod1}.
Once a loop is formed, its subsequent evolution in the Kibble
regime is also considerably different than during scaling.
In particular, the typical initial loop size as well as the
subsequent loop evolution are both strongly modified.
Both of these effects can modify the resulting dark matter density.

\subsection{Loop Production and Evolution with Friction}

  The evolution of linear perturbations on long strings and closed
string loops in the presence of friction was studied
in Refs.~\cite{Vilenkin:1991zk,Garriga:1993gj}.
In Ref.~\cite{Garriga:1993gj} it was found that linear fluctuations on a long
string of wavelength larger than $\ell_f$ are overdamped
and stretched.  For wavelengths much smaller than $\ell_f$, the damping time is
on the order of $\ell_f$, which is much longer than the typical
period of oscillation but much shorter than the Hubble time.
Hence these small fluctuations oscillate and lose energy to friction
very quickly relative to the Hubble time.  Given these results
and the picture of loop formation of Ref.~\cite{Dubath:2007mf},
we expect that the typical initial loop size during friction
is on the order of $\ell_f$.\footnote{We emphasize however that
the picture of loop formation obtained in Ref.~\cite{Dubath:2007mf}
was developed under the assumption of long string scaling,
and did not consider the effects of friction.}
Fluctuations smaller than $\ell_f$ are damped out quickly,
while those larger than $\ell_f$ grow more slowly than the long
string correlation length $L$, and therefore shrink relative to $L$.
Thus, we expect that fluctuations build up near $\ell_f$,
which in turn sets the typical size of a loop when it is formed.

  To model the evolution of string loops during friction,
we will take the results of Ref.~\cite{Garriga:1993gj} for the
evolution of a circular loop to be representative of the evolution
of general loops.  (Indeed, friction tends to make the loops more
circular.) Ref.~\cite{Garriga:1993gj} finds that loops smaller
than $\ell_f$ oscillate essentially freely and lose their energy
to the thermal background according to \beq
\left.\mu\frac{d\ell}{dt}\right|_{friction} \simeq
-\mu\,\frac{\ell}{\ell_f}. \eeq It follows that such loops lose
energy over the time scale $\ell_f$, much less than the Hubble
time in the friction regime. We do not expect these interactions
between the strings and the thermal background to be a significant
source of dark matter. 

  The motion of loops larger than $\ell_f$ is overdamped.
They evolve according to
\beq
R\dot{R}\simeq \ -\frac{\ell_f}{2a}
\label{rdamp}
\eeq
where $R$ is the comoving coordinate radius of the loop
(and $aR$ corresponds to the physical loop radius).
The solution of Eq.~\eqref{rdamp} implies that loops
of initial size smaller than the long string scale $L$
shrink down to size $\ell_f$ in less than about a
Hubble time.  Once they do, the overdamping approximation
of Eq.~\eqref{rdamp} breaks down, and the loops begin to
oscillate and decay away.

  In summary, loops are formed in the friction regime with
a typical initial length close to $\ell_f$.
The loop formation rate per unit volume per unit length is
\beq
r(\ell_i,t_i) \simeq \lrf{t_*}{t_i}\,t_i^{-4}\;
\delta(\ell_i-\ell_f(t_i)),~~~t_i<t_*.
\label{rfric}
\eeq
Any loop formed with initial length larger than $\ell_f$
will shrink down to length $\ell_f$ in less than about a Hubble time.
Loops smaller than $\ell_f$ execute underdamped oscillations,
transferring their energy to the thermal background
and decaying away over the time scale $\ell_f$.
The loop length $\ell$ in this regime evolves according to
\beq
\frac{d\ell}{dt} = -\frac{\ell}{\ell_f}
- \Gamma\,G\mu - p_c\sqrt{\frac{w}{\ell}}.
\eeq
Here, the first term comes from friction, the second from gravitational
radiation, and the third from cusping.

  Before going on to compute the dark matter density generated by
the decaying loops, let us make note of the fact that at the end
of the friction-dominated Kibble regime, the long string network
is smoothed out nearly all the way up to the Hubble scale.  In the
loop formation picture of
Ref.~\cite{Polchinski:2006ee,Polchinski:2007qc,
Rocha:2007ni,Dubath:2007mf}, small fluctuations on long strings
giving rise to loops originate from fluctuations of Hubble size
that have slowly shrunk.  Thus it will take some time for small
scale structure to build up on the long strings, and the typical
initial loop size will be initially larger than the gravitational
damping length $l_{GW}$.  We follow Ref.~\cite{Martins:1995tg} and
model this transitional period by writing the initial typical loop
size parameter as
\beq
\alpha_{eff}(t) =
\frac{2+\alpha(t/t_*)^{\xi}}{1+(t/t_*)^{\xi}},~~~~~t>t_*,
\label{alfeff}
\eeq
with the exponent $\xi \simeq 1$. A na\"ive application of the results of
Refs.~\cite{Polchinski:2006ee,Polchinski:2007qc,Rocha:2007ni,Dubath:2007mf}
suggests that $\xi=0.9$ in the radiation era.  We will study a
range of values of $\xi$.

\subsection{Dark Matter with Friction}

\begin{figure}[ttt]
\begin{center}
        \includegraphics[width = 0.7\textwidth]{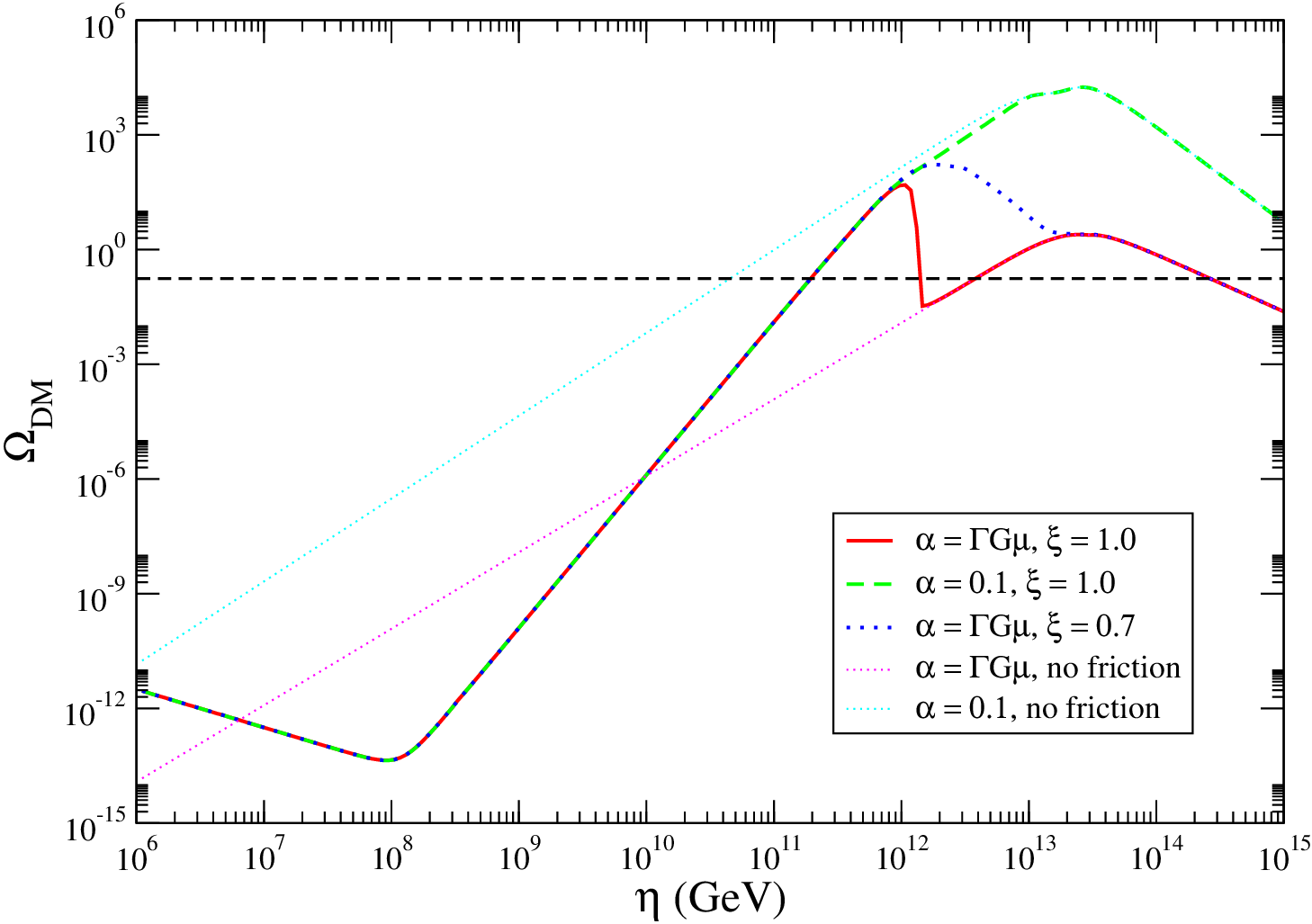}
\end{center}
\vspace{0.2cm}
\caption{Dark matter density due to loop cusping for $\epsilon_{cusp} = 1$,
$p_c=1$, and $t_{fo} = 2\times 10^{16}\,\gev^{-1}$ as a function of the
symmetry breaking VEV $\eta$ in the presence of friction.
The various curves correspond to different values of the
initial loop size parameter $\alpha$.
The black dashed horizontal line indicates the observed dark matter
relic density.
}
\label{dmfric}
\end{figure}

  The discussion above provides all the ingredients needed to evaluate
the dark matter density created by a string network in the presence 
of friction using Eq.~\eqref{mastereq1}. We do so here under 
the assumption that loop cusping is the dominant source of dark matter 
from the strings. Our results are presented in Fig.~\ref{dmfric}, where we
show the dark matter density due to cosmic strings as a function
of the symmetry breaking VEV $\eta$.  In making this plot, we have
also set $\zeta = 10$, $p_c=1$, $\epsilon_{cusp} = 1$, and we have
again taken the freeze-out time to be $t_{fo}=2\times
10^{16}\,\gev^{-1}$. 
We have also fixed an overall prefactor in
Eq.~\eqref{rfric} (equal to $2\zeta/(2\!+\!\alpha$)) to ensure 
that the loop formation rate is continuous as $t_i$ crosses $t_*$. 
The different curves in Fig.~\ref{dmfric} correspond to 
different values of the scaling
regime loop size parameter $\alpha$ and the exponent $\xi$
appearing in Eq.~\eqref{alfeff}.

  The dark matter density curves in Fig.~\ref{dmfric}
show three distinct regions, with the transitions between
them occurring around $\eta = 10^8\,\gev$ and $\eta = 10^{12}\,\gev$.
For values of $\eta$ well above $10^{12}\,\gev$, the curves
coincide with those obtained in the absence of friction.
Such large values of $\eta$ imply a value
of $t_*$ that is much smaller than the freeze-out time $t_{fo}$,
so that all loops generated while friction is relevant decay away
before $t_{fo}$.  On the other hand, when $\eta$ lies below $10^{12}\,\gev$,
the most important contribution to the dark matter comes from
loops formed while friction dominates the network evolution,
$t_i< t_*$.

  In the region $10^8\,\gev \lesssim \eta \lesssim 10^{12}\,\gev$
most of the dark matter is produced by loops that are formed in the
friction era, with $t_i < t_*$.  As a result, the dark matter
density is independent of the scaling value of $\alpha$
for this range of $\eta$.  The largest contribution to the dark matter
in this region comes from loops that are also long-lived,
with $t_i < t_{fo}$.  This enhances the amount of dark matter
formed because, with the loop distribution function of Eq.~\eqref{rfric},
the integrand of the $t_i$ integral in Eq.~\eqref{mastereq1}
is a rapidly decreasing function of $t_i$.
Increasing $\eta$ in this region leads to larger initial loop sizes
that further extend the loop lifetime. However, the minimal value of
$t_i$ for which loops decay after $t_{fo}$ decreases more slowly
with increasing $\eta$ than $t_*$. The sharp transition near $\eta
= 10^{12}\,\gev$ occurs when these two quantities become equal.
When this happens, increasing $\eta$ further decreases the initial
loop size (see Eq.~\eqref{alfeff}) for the dominant DM loops, and
these loops go quickly from being long-lived to being short-lived,
with the dominant loops being formed near $t_i = t_{fo} > t_*$.
The transition is more gradual when the exponent $\xi$ in
Eq.~\eqref{alfeff} is less than $\xi=1.0$, which can be seen
by comparing with the curve for $\xi = 0.7$.

  For $\eta \lesssim 10^{8}\,\gev$,
the initial loop size $\ell_f$ becomes small enough
that the loops are short-lived, decaying away within a Hubble time.
Thus, the majority of the dark matter produced for these smaller
values of $\eta$ comes from loops formed near the freeze-out time,
$t_i\simeq t_{fo}$, with initial size close to $\ell_{f}(t_{fo})$.
As $\eta$ decreases below $10^8\,\gev$, we find that a larger
fraction of the energy of each loop is lost to cusping,
thereby increasing the dark matter density.  Even so,
the dark matter relic density generated by cosmic strings 
for $\eta \sim 10^3\,\gev$ is a negligibly small fraction of 
the observed value.

\section{Conclusions\label{conclusion}}

  We have investigated the cold dark matter density
created by the decays of loops of gauge cosmic strings.
Dark matter is produced by string loops when they form
cusps.  At a cusp, a small portion of the string loop annihilates into
its constituent fields, which can then cascade down to lighter states
such as dark matter particles.  Our results provide constraints
on extensions of the gauge symmetry group of the standard model
that give rise to cosmic strings in the early universe.
The string loops that give decay to dark matter are themselves
created continually by the network of long, horizon-length
strings. We have studied the amount of dark matter generated when
this network is in the \emph{scaling} regime, and when its
evolution is dominated by frictional forces. Both cases are
physically relevant as the presence or absence of significant
frictional interactions depends on the details of the symmetry
breaking from which the strings originate. 

  The dark matter density generated by a string network
in the scaling regime can potentially be enough to explain the 
observed relic density when the breaking VEV $\eta$ 
exceeds $10^{10}\,\gev$.  The amount of dark matter produced by 
this mechanism is greatest when the initial loop size approaches 
a significant fraction of the cosmological horizon, 
although much smaller initial loop sizes can also generate
sizeable densities of dark matter.  However, larger loops
have cusp decay products that are highly boosted, and this can
reduce the fraction of the loop energy that becomes cold dark matter.
For values of $\eta$ below $10^{10}\,\gev$, the amount cold 
dark matter created by a scaling string network is always well below 
the observed value.  

  String networks that are strongly influenced by frictional 
interactions can also generate a dark matter density that is
equal to or greater than the observed value.  This can occur 
if the symmetry breaking scale is greater than about $10^{11}\,\gev$.  
For very large values of the symmetry breaking scale, 
above about $10^{13}\,\gev$, frictional effects decouple 
well before the non-thermal dark matter is created; the frictional
interactions do not alter the resulting dark matter density.
When the symmetry breaking scale is much below $10^{11}\,\gev$, 
the contributions to the dark matter density in the presence of
friction are very small.  This contrasts with and is less constraining 
than the result of Ref.~\cite{Jeannerot:1999yn}, in which the effects of
cusping on the decays of loops were not included in computing the 
dark matter density.

  We conclude from our investigation that the non-thermal cold
dark matter relic density is safely below the observed value
when the symmetry breaking scale is less than $\eta \simeq 10^{10}\,\gev$.
For larger symmetry breaking scales, the non-thermal dark matter
density produced by decaying string loops can potentially be as large
or larger than the observed value.  Thus, our results also put 
constraints on new gauge symmetries broken at very high scales 
based on the requirement that they do not overproduce dark matter.  
These constraints are model-dependent, in that they depend on the 
effective branching fraction of the decaying string
loops into dark matter, but they can be more severe
than the constraints from the more traditional cosmic string signatures
such as gravitational radiation and gravitational 
lensing~\cite{Hindmarsh:1994re,sbook}.  
For gauge symmetries broken at lower scales that could be probed by 
the LHC, we find that the contribution to the dark matter density 
from their corresponding scaling string loops is negligibly small.


\section*{Acknowledgements}

  We thank Aaron Pierce, Joseph Polchinski, Jorge Rocha,
and James Wells for helpful discussions and comments.
We also thank Aaron Pierce and James Wells for suggestions and comments
on the manuscript, and Nima Arkani-Hamed for pushing us to look more
closely at the effect of large boosts from string cusps.
This work was supported by the DOE,
the Michigan Center for Theoretical Physics~(MCTP),
and the National Science Foundation under Grant No.~PHY05-51164.
The authors would also like to
thank the Kavli Institute for Theoretical Physics for their
hospitality while this work was being completed.



\end{document}